\documentclass[epj,twocolumn]{webofc}
\usepackage[varg]{txfonts}   
\woctitle{Hadron Collider Physics Symposium 2012}
\begin{document}
\title{Backgrounds in $H\rightarrow WW^{(\ast)} \rightarrow \ell\nu\ell\nu$
  with ATLAS}
%
%

\author{Tomo Lazovich\inst{1}\fnsep\thanks{\email{tlazovich@physics.harvard.edu}} On behalf of the ATLAS Collaboration 
}

\institute{Harvard University, Department of Physics, 17 Oxford St., Cambridge, MA 02138, USA}

\abstract{%
  We present techniques used to estimate the backgrounds in the search for the Standard Model Higgs boson in the $H\rightarrow WW^{(\ast)} \rightarrow \ell\nu\ell\nu$ 
decay channel with the ATLAS experiment at the LHC. The dataset corresponds to 13 $\text{fb}^{-1}$ of integrated luminosity taken at a center of mass energy of 8 TeV. Only the final states with an electron, muon, and zero or one jet are presented here. 
}
\maketitle
\section{Introduction}
\label{intro}

In July 2012 the ATLAS \cite{ATLASJinst} and CMS \cite{CMSJinst}
experiments at the LHC announced the discovery of a new particle
consistent with the long-sought Higgs boson \cite{ATLASDiscovery,
  CMSDiscovery}. The results presented here constitute an update of
the $H\rightarrow WW^{(\ast)}\rightarrow \ell\nu\ell\nu$ analysis with
a dataset of 13 $\text{fb}^{-1}$ taken at a center of mass energy of 8
TeV \cite{HCPConfNote}. In particular, we summarize the methods of background estimation for this search channel, which focuses on the low mass Higgs signal region. 

The $WW^{(\ast)} \rightarrow \ell\nu\ell\nu$ decay channel of the Higgs boson has a final state defined by two leptons and missing transverse energy ($E_{T}^{\text{miss}}$) from neutrinos which escape detection. The analysis presented here considers only the final states with one electron, one muon, and zero or one jets with transverse momentum ($p_{T}$) greater than 25 GeV. The leptons are required to be isolated and the leading (subleading) lepton must have $p_{T} > 25\text{ } (15)$ GeV. Additionally, the event must have relative missing transverse energy ($E_{T, \text{rel}}^{\text{miss}}$) greater than 25 GeV, where $E_{T, \text{rel}}^{\text{miss}} = E_{T}^{\text{miss}}\sin(\text{min}(\Delta\phi, \frac{\pi}{2}))$ and $\Delta\phi$ is the azimuthal angle between the $E_{T}^{\text{miss}}$ and the nearest reconstructed lepton or jet. This definition helps to reject events where a mismeasurement of one of the reconstructed objects is a major source of the $E_{T}^{\text{miss}}$.  After these pre-selection cuts, the signal region is divided into zero and one jet bins and additional topological cuts (specific to each bin) are applied to discriminate the Higgs signal from background contributions. 

\begin{figure}[h!]
\centering
\includegraphics[width=6cm,clip]{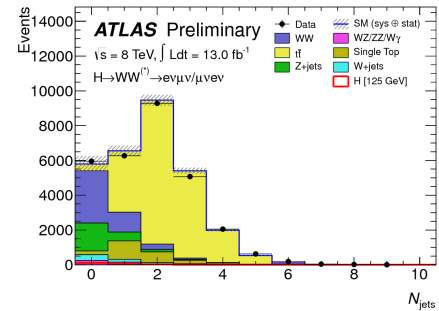}
\caption{The jet multiplicity distribution after signal pre-selection cuts are applied \cite{HCPConfNote}.}
\label{fig-jetmult}       
\end{figure}

Many processes in the Standard Model (SM) produce final states similar to that in $H\rightarrow WW^{(\ast)} \rightarrow \ell\nu\ell\nu$. The largest background contribution is the irreducible SM WW background. The next largest background contributions come from $t\bar{t}$ and single top production. These backgrounds are primarily relevant for larger jet multiplicity bins but are also present in the zero jet bin. Another important background for the analysis is the W+jets background. Here a single W boson is produced in association with one or more jets, and one of the jets fakes a final state lepton. Other backgrounds which will not be discussed in great detail include the Z+jets and diboson ($WZ$, $ZZ$, $W\gamma$) backgrounds. Figure~\ref{fig-jetmult} shows the jet multiplicity distribution before the selection separates the events into jet multiplicity bins. After the pre-selection, the WW background is the dominant background in the zero jet bin while the top background dominates the higher jet multiplicity bins. 

\section{W+jets and Other Minor Backgrounds}

The W+jets background arises from SM W boson production in association with jets where one jet produces an object reconstructed as a lepton. This can be a real lepton produced by heavy quark decay or a product of the jet fragmentation that is incorrectly reconstructed as an electron. The W+jets background contribution is estimated by a data-driven method called the ``fake factor'' method. First, a control region is defined in data by requiring one lepton with the same identification and isolation criteria as the signal leptons. The second lepton is required to be anti-identified, satisfying loosened isolation criteria and failing at least one identification requirement. These events are then required to pass the full signal selection.

A fake factor, the ratio of the number of lepton candidates passing all identification requirements and signal selections to the number that are anti-identified, is derived in an inclusive data dijet sample. This factor is used to scale the number of events in the control region to the signal region. The total relative uncertainty on the estimate is 50\%, dominated by the systematic uncertainty on the fake factor. 

\begin{figure}[h!]
\centering
\includegraphics[width=6cm,clip]{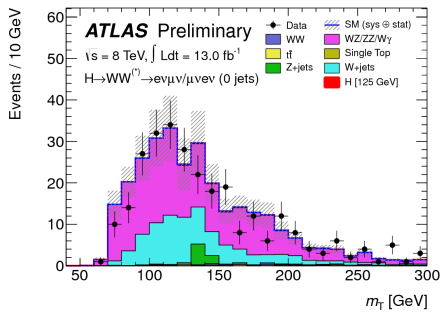}
\caption{$m_{T}$ in the zero jet same sign validation region \cite{HCPConfNote}.}
\label{fig-wjets}      
\end{figure}

Figure~\ref{fig-wjets} shows the transverse mass ($m_{T}$) in a same sign validation region (where two same sign rather than two opposite sign leptons are required). This region is composed largely of W+jets and $WZ/ZZ/W\gamma$ backgrounds and is used to validate the modeling of kinematic variables for these samples. This region shows that the $m_{T}$ is well modeled (within statistics) for these samples. 

Here we briefly mention other minor backgrounds which will not be discussed in further detail.
First, the Z+jets background comes from a case
where the Z decays to two leptons and there is fake missing transverse energy in the event due to the calorimeter
resolution. This background is normalized to data in a control region
requiring $m_{\ell\ell} < 80$ GeV and $\Delta \phi_{\ell\ell} > 2.9$. Finally, the normalizations for the remaining backgrounds ($WZ/ZZ/W\gamma$) are taken from Monte Carlo simulation.

\begin{figure}[h!]
\centering
\includegraphics[width=6cm,clip]{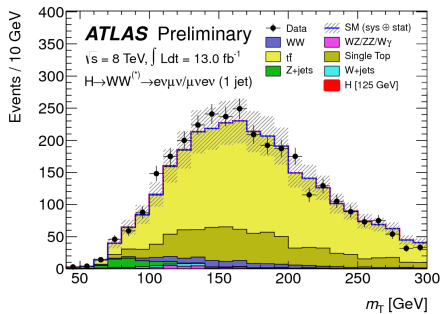}
\caption{$m_{T}$ distribution in the top one jet control region. The rate predicted by Monte Carlo simulation has not yet been normalized to the data. \cite{HCPConfNote}}
\label{Top1jMT}      
\end{figure}

\section{$t\bar{t}$ and Single Top Backgrounds}
\label{sec-ttbar}

The $t\bar{t}$ and single top backgrounds are normalized together in control regions (CR) separated by jet multiplicity. The one jet bin is normalized in a CR defined with the same pre-selection as the signal region (SR) and at least one b-tagged jet. For the zero jet bin, there are two CRs used for the background estimate. First, a CR with only the SR pre-selection is used to estimate the fraction of top events passing a jet veto. A second, b-tagged CR is then used to estimate the probability of having no other jets reconstructed in the event and is used as a correction to the fraction estimate from the first CR. 

Figure~\ref{Top1jMT} shows the $m_{T}$ in the one jet CR before any normalization factors are applied. The normalization factors (NF), or ratio between the data and Monte Carlo predictions, derived via these methods are $1.04 \pm 0.05$ (stat.) for the zero jet channel and $1.03 \pm 0.02$ (stat.) for the one jet channel.

\section{Standard Model WW Background}
\label{sec-WW}

The SM WW background is estimated in a CR which uses the SR pre-selection cuts (two leptons, missing transverse energy) and is separated into jet bins. While the SR requires $m_{\ell\ell} < 50$ GeV, the WW CR requires $m_{\ell\ell} > 80$ GeV. This is the largest background in the signal region. The WW modeling in simulation is done with a tune of Powheg for event generation and Pythia 8 for parton showering. 

\begin{figure}[h!]
\centering
\includegraphics[width=6cm,clip]{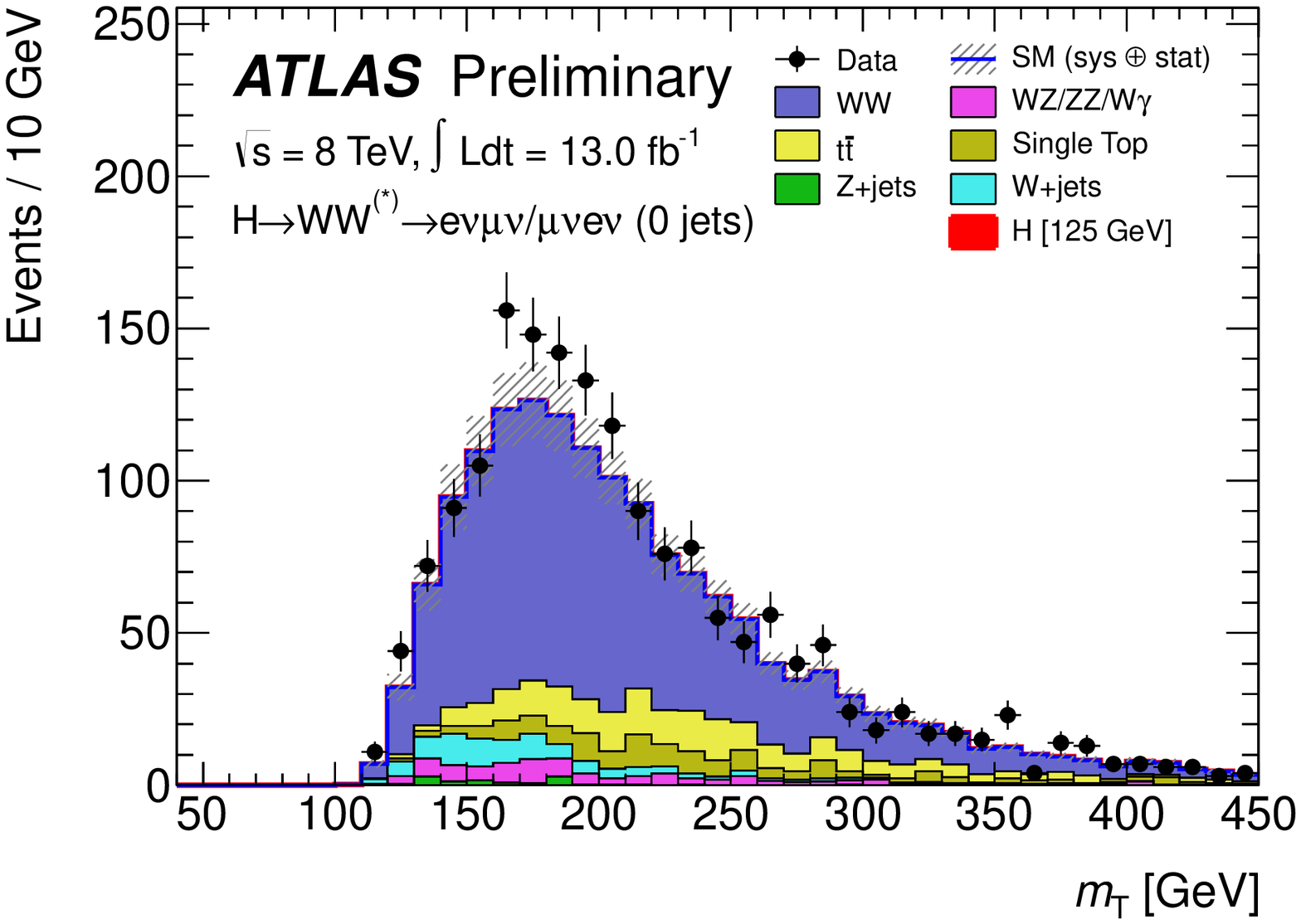}\\
\includegraphics[width=6cm,clip]{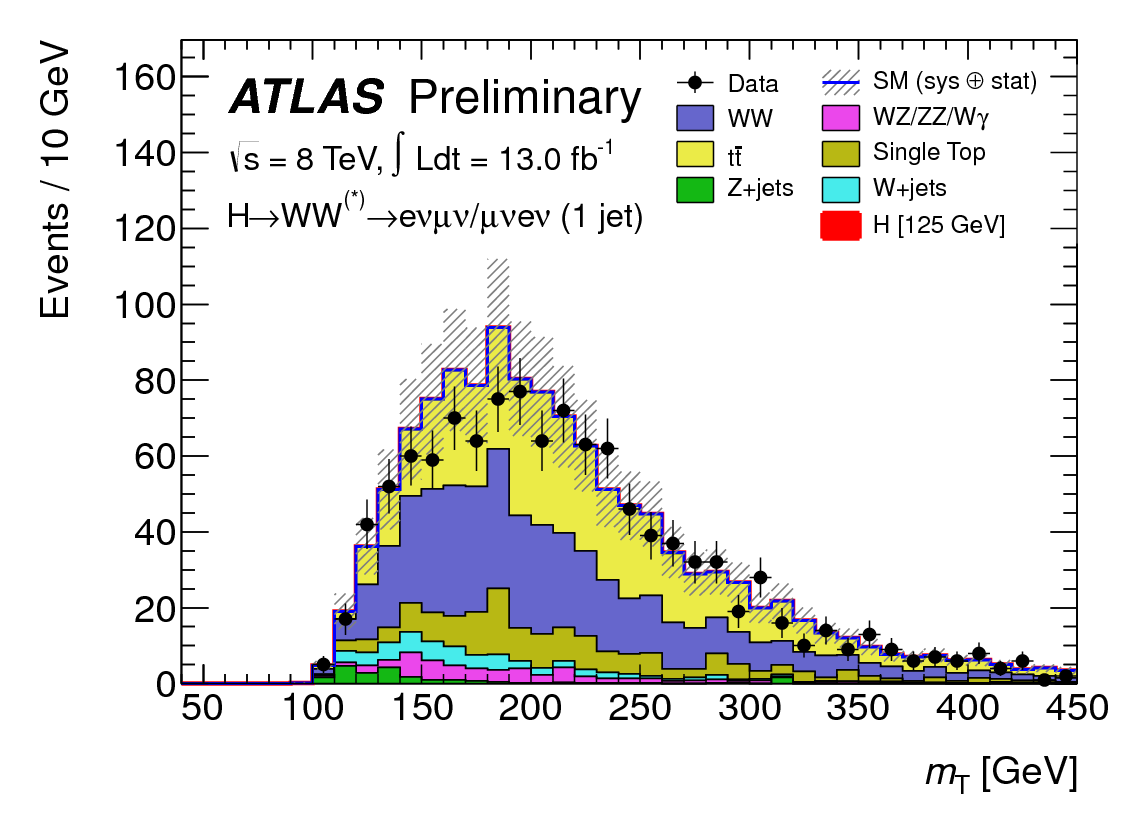}\\
\caption{The $m_{T}$ distribution in the WW zero (one) jet CR is shown in the top (bottom) plot. The WW rate predicted by Monte Carlo simulation has not yet been normalized to the data. The top backgrounds have been normalized according to the procedure described in the text \cite{HCPConfNote}.}
\label{WW0jMT}       
\end{figure}

Figure~\ref{WW0jMT} shows the $m_{T}$ in the WW zero and one jet CR, before the application of any WW NF. In both the zero and one jet (but particularly in the one jet) WW CRs, there is a non-negligible contrbution from the top backgrounds. Therefore, the top backgrounds are first normalized using the procedures described in Section~\ref{sec-ttbar} before all of the non-WW backgrounds are subtracted from the event yields in the CR to derive the final normalization. The ratio of the data (with non WW background subtracted) to the WW simulation prediction is $1.13 \pm 0.04$ (stat.) in the zero jet CR and $0.84\pm 0.08$ (stat.) in the one jet CR. The NF differ between the zero and one jet channels because these CR are correcting for the over-prediction of the jet multiplicity distribution by the current Powheg+Pythia 8 tune used by ATLAS.

\begin{table}
\centering
\resizebox{8cm}{!}{
\begin{tabular}{c| c c c c |c}
\hline\hline
Background & Stat. (\%) & Theory (\%) & Expt. (\%) & Crosstalk (\%) & Total (\%)  \\\hline
$WW$, $H + \text{0-jet}$ & 3.3 & 7.2 & 1.5 & 6.2 & 13 \\
$WW$, $H + \text{1-jet}$ & 9 & 8 & 12 & 34 & 54 \\
top, $H + \text{1-jet}$ & 2 & 8 & 29 & 1 & 37 \\\hline\hline
\end{tabular}
}
\caption{Total uncertainties on backgrounds normalized using simple NF scaling in CRs \cite{HCPConfNote}.}
\label{tab-uncert}       
\end{table}

\section{Background Predictions and Uncertainties}

Table~\ref{tab-uncert} shows the total uncertainties on the background normalization for the backgrounds which use the simple data to simulation scaling in the CR for normalization. Theoretical uncertainties on the estimates include differences due to the choice of generator and parton shower/underlying event as well as other contributions. The experimental uncertainties are dominated the jet energy scale and resolution and, in the one jet bin, the $b$ tagging efficiency. The ``Crosstalk'' column refers to uncertainties on other backgrounds which must be subtracted from the CR before the normalization of the desired background can be computed. Notice in particular that the WW one jet normalization has a large contribution from crosstalk due to the fact that a top background contribution must be subtracted from the CR before the normalization is computed.

Table~\ref{tab-nf} shows the NF derived for all of the backgrounds whose normalizations are taken from data. In the case of everything except the top zero jet background, this factor is simply the ratio of the number of Monte Carlo events to data events in the appropriate CR for that background. We can see that most of the backgrounds do not require very large corrections to their normalization (none more than 16\%). 

\begin{table}
\centering
\begin{tabular}{c| c c c c |c}
\hline\hline
Background & 0 jet NF & 1 jet NF  \\\hline
Top & $1.04 \pm 0.05$ & $1.03 \pm 0.02$ \\
WW & $1.13 \pm 0.04$ & $0.84 \pm 0.08$ \\
Z+jets & $0.87\pm 0.03$ & $0.85 \pm 0.03$ \\\hline\hline
\end{tabular}
\caption{Normalization factors (NF) for all backgrounds whose normalizations are taken from data \cite{HCPConfNote}.}
\label{tab-nf}       
\end{table}

\begin{figure}[h!]
\centering
\includegraphics[width=6cm,clip]{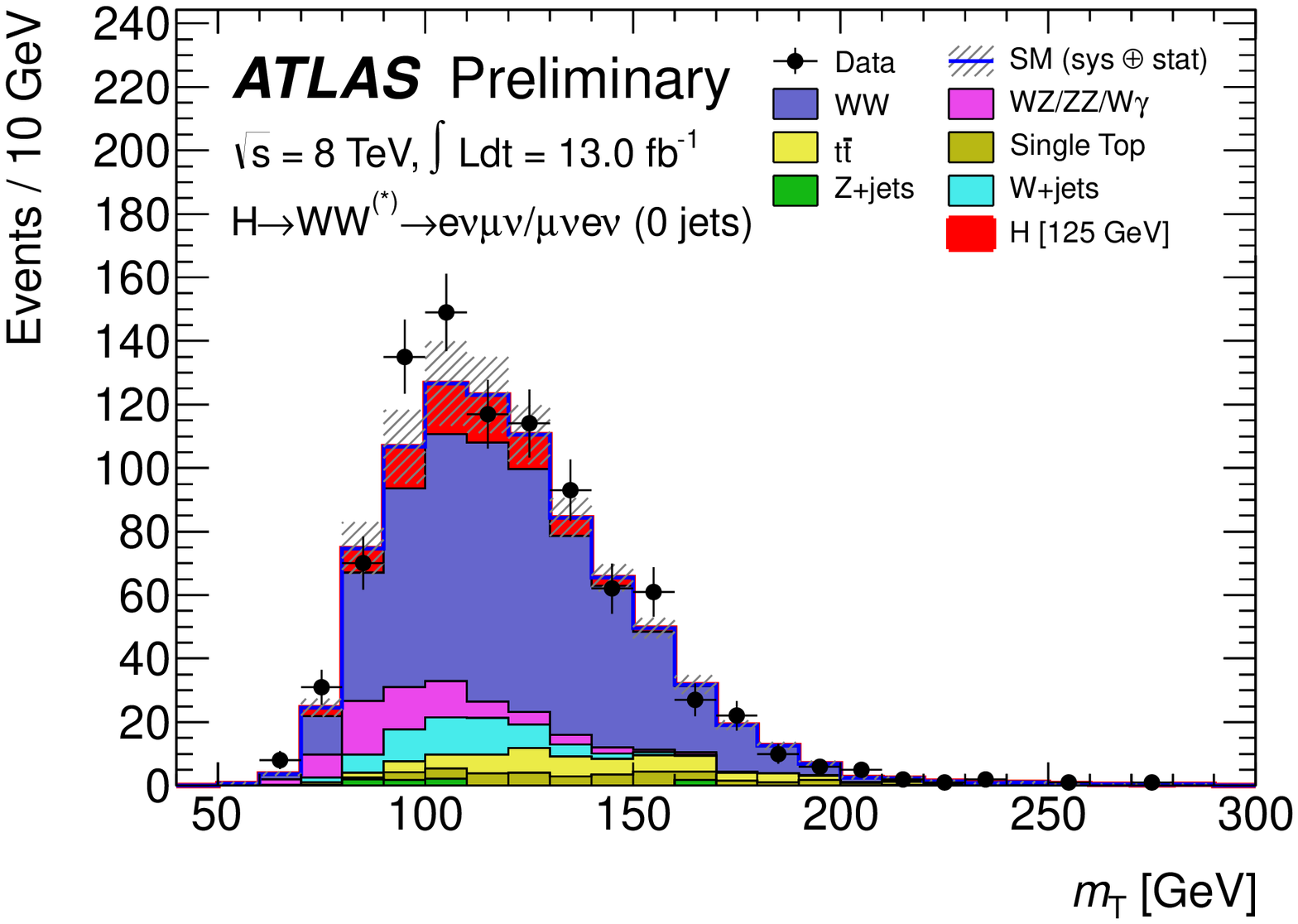}
\includegraphics[width=6cm,clip]{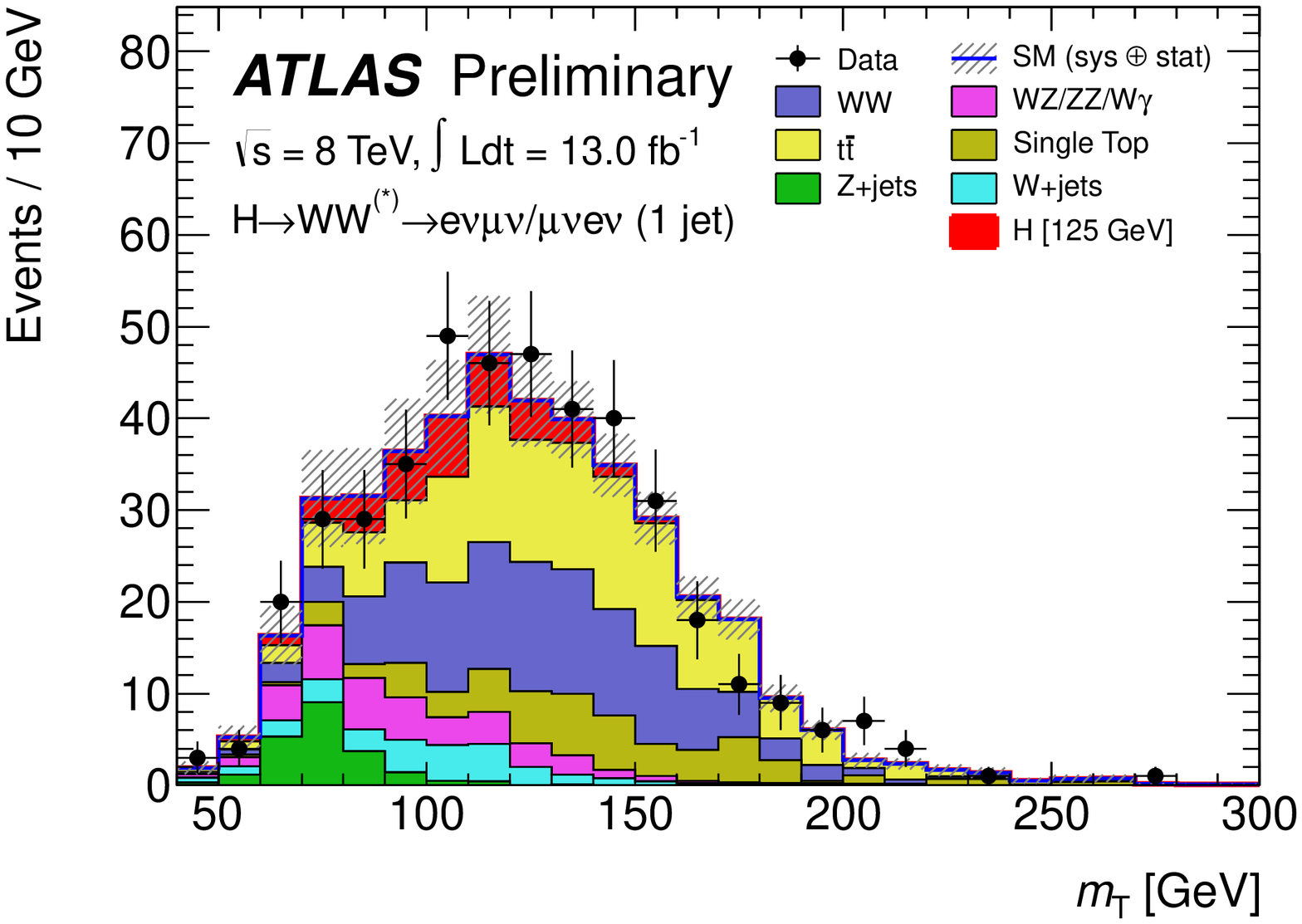}
\caption{$m_{T}$ distribution in the zero (one) jet bin after all signal selection cuts in the top (bottom) plot \cite{HCPConfNote}. The expectation for a 125 GeV Higgs signal is shown in red.}
\label{fig-Sig0jMT}       
\end{figure}

Figure~\ref{fig-Sig0jMT} shows the $m_{T}$ distribution after the signal selection cuts have been applied for the zero and one jet bin. It can be seen here that the WW background is dominant in the zero jet bin, while both WW and top are dominant in the one jet bin. In zero jet, there is a total of $774 \pm 9$ (stat.) expected background events, and $555 \pm 5$ (stat.) of those are SM WW events. In the one jet, out of $386 \pm 5$ (stat.) total expected background events, $118 \pm 2$ (stat.) are SM WW events while $134 \pm 5$ (stat.) are $t\bar{t}$ events. The difference between the background expectation and the 917 (433) events observed in data in the zero (one) jet bin are due to the presence of the Higgs-like signal.

\section{Conclusion}

A wide array of estimation methods can be employed to understand the complicated background processes that factor into a search for a $H\rightarrow WW^{(\ast)} \rightarrow \ell\nu\ell\nu$ signal. Simple data to simulation scaling in control regions is used for backgrounds such as SM WW (or top in the one jet bin) where the variable shapes are well modeled but their normalizations may be incorrect. More complicated data-driven methods, such as the W+jets fake factor method, can also be used when the backgrounds are not well modeled by simulation alone. 

%

\vspace{-20pt}
\end{document}